\documentclass{article}
\PassOptionsToPackage{numbers, compress}{natbib}

\usepackage[preprint]{neurips_2023}

\usepackage[utf8]{inputenc} 
\usepackage[T1]{fontenc}    
\usepackage{hyperref}       
\usepackage{url}            
\usepackage{booktabs}       
\usepackage{amsfonts}       
\usepackage{nicefrac}       
\usepackage{microtype}      
\usepackage{xcolor}         
\usepackage{amsmath}
\usepackage{bm}
\usepackage{marvosym}
\usepackage{xspace}
\usepackage{xcolor}
\usepackage{soul}
\usepackage[framemethod=tikz]{mdframed}
\usepackage{enumitem}
\usepackage{wrapfig}
\usepackage{caption}
\usepackage{multirow}
\usepackage{subfigure}
\usepackage{adjustbox} 

\newcommand{\ie}{\emph{i.e.,}\xspace}
\newcommand{\eg}{\emph{e.g.,}\xspace}

\newcommand{\paratitle}[1]{\vspace{1.5ex}\noindent\textbf{#1}}

\newcommand{\ignore}[1]{}

\definecolor{condition}{HTML}{E0EBF6}

\definecolor{candidate}{HTML}{E5EFDB}

\definecolor{ginger}{rgb}{0.69, 0.4, 0.0}
\definecolor{pastelorange}{rgb}{1.0, 0.7, 0.28}
\definecolor{teal}{rgb}{0.0, 0.5, 0.5}
\definecolor{thistle}{rgb}{0.85, 0.75, 0.85}

\title{Slow Thinking for Sequential Recommendation}

\renewcommand\footnotemark{}

\author{%
Junjie Zhang\textsuperscript{\textmd{1}} \And Beichen Zhang\textsuperscript{\textmd{1}} \And Wenqi Sun\textsuperscript{\textmd{1}} \And Hongyu Lu\textsuperscript{\textmd{2}} \AND Wayne Xin Zhao\textsuperscript{\textmd{1}\Letter} \qquad Yu Chen\textsuperscript{\textmd{2}} \qquad Ji-Rong Wen\textsuperscript{\textmd{1}}
  \thanks{\Letter\ Corresponding author.} \\ \\
  \textsuperscript{1}Gaoling School of Artificial Intelligence, Renmin University of China \\
  \textsuperscript{2}WeChat, Tencent \\
  \small\texttt{junjie.zhang@ruc.edu.cn\quad batmanfly@gmail.com}
}

\begin{document}

\maketitle

\begin{abstract}
To develop effective sequential recommender systems, numerous methods have been proposed to model historical user behaviors. Despite the effectiveness, these methods share the same \textbf{fast thinking} paradigm. That is, for making recommendations, these methods typically encodes user historical interactions to obtain user representations and directly match these representations with candidate item representations. However, due to the limited capacity of traditional lightweight recommendation models, this one-step inference paradigm often leads to suboptimal performance. To tackle this issue, we present a novel \textbf{slow thinking} recommendation model, named \textbf{STREAM-Rec}. Our approach is capable of analyzing historical user behavior, generating a multi-step, deliberative reasoning process, and ultimately delivering personalized recommendations. In particular, we focus on two key challenges: (1) identifying the suitable reasoning patterns in recommender systems, and (2) exploring how to effectively stimulate the reasoning capabilities of traditional recommenders. To this end, we introduce a three-stage training framework. In the first stage, the model is pretrained on large-scale user behavior data to learn behavior patterns and capture long-range dependencies. In the second stage, we design an iterative inference algorithm to annotate suitable reasoning traces by progressively refining the model’s predictions. This annotated data is then used to fine-tune the model. Finally, in the third stage, we apply reinforcement learning to further enhance the model’s generalization ability. Extensive experiments validate the effectiveness of our proposed method.

\end{abstract}

\section{Introduction}

Personalized recommendation systems are essential for helping users navigate the online content, significantly improving user experiences across platforms such as e-commerce, streaming services, and news portals. Among the various recommendation approaches, \emph{sequential recommendation} has gained considerable attention for its effectiveness in modeling the temporal sequence of user behaviors and capturing the evolving nature of user preferences~\cite{S3Rec,SASRec}. Especially, in the literature of sequential recommendation, many impressive methods have been proposed to improve the recommendation performance (\eg RNNs~\cite{gru4rec}, CNNs~\cite{caser}, and Transformers~\cite{BERT4Rec}). 

Although the specific techniques are different, most existing methods share the same \textbf{fast thinking} paradigm. Typically, they use a sequential model to encode user historical behaviors. The resulting user representations are then \textbf{directly} used to ``match'' candidate item representations. Items with the highest matching scores are selected for recommendation. As a result, the recommendation performance of these models largely depends on their ability to learn effective representations. This undoubtedly poses a significant challenge for small recommendation models.

Despite recent advancements, accurately predicting the next item remains a significant challenge, especially when user intentions are subtle or evolve rapidly. In such cases, it is often unclear how to infer the target item from sparse interaction data using the simple matching method. To address this issue, a number of researches have explored leveraging the reasoning capabilities of large language models (LLMs) to analyze user behavior and generate high-quality representations, thereby enhancing traditional lightweight recommendation models (LLM4Rec)~\cite{wu2023llm4rec_survey_1, lin2023_llm4rec_survey_2, fan2023_llm4rec_survey_3, zhang_2024_agentcf_www}. However, these approaches also bring new challenges, such as potential inconsistencies between the recommendation domain-specific knowledge and the universal world knowledge embedded in LLMs, as well as concerns regarding the efficiency of reasoning processes within LLMs~\cite{zhang_2023_instructrec_arxiv, bao_2023_tallrec_recsys}. These issues lead us to a central question: \textbf{Can we enhance recommendation performance by stimulating the reasoning capabilities of traditional recommender systems?} More recently, increasing evidence shows that \textbf{test-time scaling} endows the model with stronger logical reasoning ability, enabling more powerful generation~\cite{tang_2025_rearec_arxiv, zhang_2025_test_time_scaling_survey_arxiv, min_2024_still_v2_arxiv}.

Motivated by recent advances in slow thinking~\cite{chen_2025_still_v1_arxiv}, we introduce a novel sequential recommendation approach aimed at generating more appropriate recommendations. The key idea is to empower the recommender system to gradually construct intermediate reasoning steps, ultimately leading to accurate predictions. While prior work in large language models has demonstrated the potential of slow thinking~\cite{zhang_2025_test_time_scaling_survey_arxiv}, there are still several key challenges to be solved. First, unlike the clearly defined logical steps used in solving math or code problems~\cite{jiang_2024_still_v1_arxiv}, it is non-trivial to formulate a reasoning process that effectively guides the model toward generating target items from users’ historical interactions. Second, traditional recommender systems are not inherently designed for slow thinking. The classical training task (\eg next item prediction~\cite{RecBole}) does not naturally encourage a multi-step, deliberative reasoning process.

To address these issues, in this paper, we propose the \textbf{S}low \textbf{T}hinking \textbf{RE}commend\textbf{A}tion \textbf{M}odel, called~\textbf{STREAM-Rec}. Given user historical interactions, our model gradually generates valuable reasoning processes, and finally arrives at suitable recommendations. To achieve this, we design a three-stage training framework, guiding the sequential recommender to follow a slow thinking paradigm. Specifically, in the first stage, the model is pre-trained on user behavior sequences to capture sequential patterns and support long-range dependency modeling.  In the second stage, we propose a novel iterative inference method for generating recommendation-specific slow thinking data. Especially, the model is guided to iteratively refine its predictions by computing and fitting the residual between its current output and the target behavior representation. These intermediate residual data serve as reasoning paths and are then collected to supervised fine-tune (SFT) our model, thereby encouraging step-by-step adjustments based on its internal state. Finally, in the third stage, we employ reinforcement learning (RL) to further enhance the reasoning and generalization abilities. The discrepancy between model predictions and target user behaviors serves as a reward signal, encouraging the model to explore more effective reasoning strategies.

To evaluate the proposed Stream-Rec, we conduct extensive experiments on real-world datasets. The results demonstrate that our approach effectively leverages the slow thinking paradigm to enhance recommendation performance, gradually generating reasoning processes and delivering more appropriate recommendations. The main contributions of this work are as follows:

\begin{itemize}
    \item We find that sequential recommenders can perform a slow thinking paradigm, generating valuable reasoning processes to support final recommendations.

    \item We design a three-stage training framework (\ie pretraining, supervised fine-tuning, and reinforcement learning) that enhances the reasoning ability of sequential recommenders.

    \item Extensive experiments demonstrate the effectiveness of
our approach in providing personalized recommendations.
\end{itemize}

\section{Methodology}

In this section, we present the proposed \textbf{S}low \textbf{T}hinking \textbf{RE}commend\textbf{A}tion \textbf{M}odel, named as~\textbf{STREAM-Rec}. 
Given historical user behavioral sequences, the model is trained to generate valuable reasoning processes that lead to more accurate recommendations.

\subsection{Preliminaries}
\label{sec:preliminary}
To ease the understanding of our approach, we first describe the traditional recommendation setting, and then introduce our task setting and the overview of our approach.

\paratitle{Traditional Matching Recommender.} Recommender systems consist of a set of users $\mathcal{U} = {u}$, a set of items $\mathcal{I} = {i}$, and a collection of interaction records between them, denoted as $\mathcal{D} = {\langle u, i \rangle}$. In this paper, we focus on the task of sequential recommendation. A user's historical interactions are represented as an ordered sequence $S_u = \{i_1, i_2, \cdots, i_n\}$, sorted by interaction time. Given user behavior sequence, traditional sequential recommender systems typically follow a direct \emph{matching} paradigm. That is, they train models to capture user preferences by encoding the behavioral sequences into user representations. These representations are then used to rank items and provide recommendations by calculating similarities with candidate item representations. Formally, the recommendation task can be formulated as predicting the next item $\hat{i}_{n+1}$ given the interaction history: 
\begin{align}
P(\hat{i}_{n+1}|{i_1, \cdots, i_n}).
\end{align}

\paratitle{Traditional Generative Recommender.} Recently, some studies propose processing user behavior sequences in a generative paradigm, thereby achieving fine-grained behavior modeling~\cite{Rajput_2023_tiger_nips, Wang_2024_letter_cikm,liu_2024_etegrec_arxiv}. Specifically, given item related representations (\eg modal embeddings), we can train a tokenizer that maps each item's representation into a sequence of semantic codes, denoted as $C_j = (c_1, c_2, \cdots, c_m)$. These codes $c$ are drawn from a shared vocabulary $\mathbb{C}$. As a result, the user behavior sequence $S_u$ can be tokenized into a sequence of tokens $\{c_1, c_2, \cdots, c_p\}$, where $p$ is the total number of tokens in the behavioral sequence. Typically, $p$ is equal to the behavior sequence length $n$ multiplied by the tokenizer length $m$.  A generative model (\eg Transformer~\cite{Raffel_2020_t5_jmlr}) is trained to autoregressively generate future tokens (\ie user target behavior $C_{i_{n+1}} = \{c_{p+1}, \cdots, c_{p+m+1}\}$), modeling the distribution:
\begin{align}
P(\hat{i}_{n+1}|S_u) = P(\{c_{p+1}, \cdots, c_{p+m+1}\}|\{c_1, c_2, \cdots, c_p\}),    
\end{align}
where $\hat{i}_{n+1}$ is parsed from the generated tokens during inference.

\paratitle{Our Proposed Slow Thinking Sequential Recommender.} In this work, we follow the classic generative recommendation paradigm, while extending the autoregressive generation process. Specifically, instead of directly predicting the next tokens, our model introduces an intermediate reasoning token sequence $O_u = \{o_1, o_2, \cdots, o_l\}$, which serves as ``intermediate decision steps''. The final recommendation is then generated conditioned on both the original user behavior and this reasoning sequence. This approach allows the model to perform a deeper analysis of user interests and provide more tailored recommendations. The overall generation process can be formalized as follows: 
\begin{align}
P(\hat{i}_{n+1}, O_u \mid S_u) = P(O_u \mid S_u) \cdot P(\hat{i}_{n+1} \mid O_u, S_u).
\end{align}

\begin{figure*}[h]
	\centering
	\includegraphics[width=0.62\linewidth]{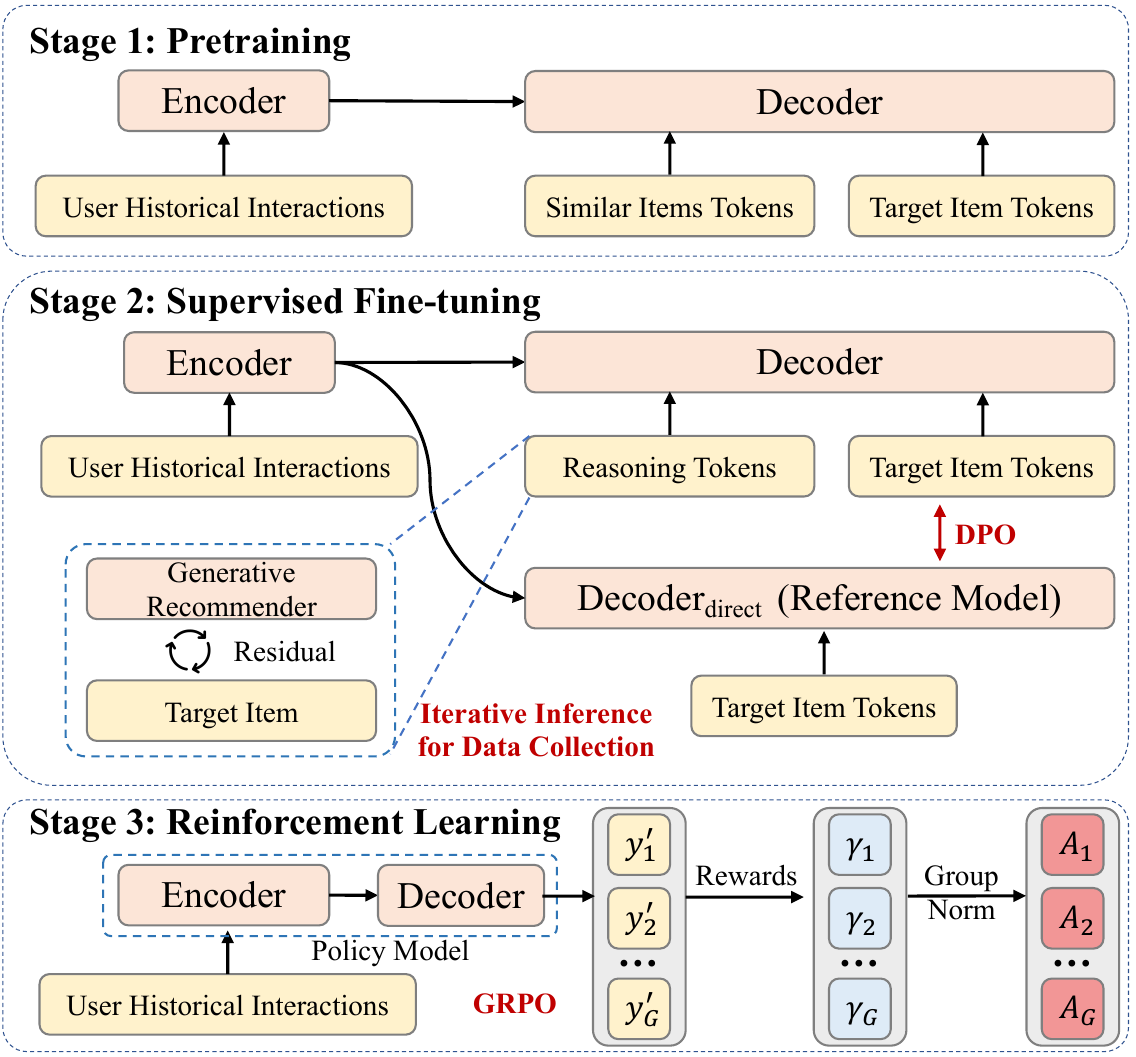}
        \caption{The overall framework of STREAM-Rec. }
	\label{fig:overall}
    \vspace{-10pt}
\end{figure*}

\paratitle{Key Challenges and Solutions.} In the literature on LLMs, two main approaches are used to enable models to follow long-chain reasoning patterns: (1) fine-tuning models on large-scale reasoning data distilled from strong reasoning models~\cite{min_2024_still_v2_arxiv}, or (2)  applying RL to guide cold-start, short-chain models to explore longer reasoning trajectories through extensive rollouts~\cite{chen_2025_still_v1_arxiv}. Nevertheless, it is non-trivial to extend these methods to the field of sequential recommender systems. First, it remains unclear which types of reasoning patterns are beneficial for recommendation, and we also lack a strong teacher model for effective distillation. Second, RL requires the model to possess a certain degree of initial competence. However, we empirically find that simply applying traditional one-step recommenders for rollout often leads to unsatisfactory results. Therefore, to induce the slow thinking reasoning capacity of traditional recommenders, we design a \textbf{three-stage training framework}. In Stage 1, we pretrain the model on large-scale user interaction data to learn general behavioral patterns and capture long-range dependencies (Section~\ref{sec:pretrain}). In Stage 2, we collect a slow thinking dataset by iteratively computing and fitting the residuals between the model’s predictions and user target behaviors. The model is then tuned on this data to capture slow thinking patterns (Section~\ref{sec:sft}). In Stage 3, we apply reinforcement learning to further improve the model’s generalization capabilities (Section~\ref{sec:rl}). The overall framework of the proposed STREAM-Rec is depicted in Figure~\ref{fig:overall}.

\subsection{Pretraining for Long-range Dependicies Modeling} \label{sec:pretrain}
In this section, we describe the first stage of our framework, \ie pretraining the STREAM-Rec.

\subsubsection{User Behavior Tokenization}
The first step in training a generative recommendation model is to tokenize user behavior data into discrete tokens. Following prior work~\cite{Rajput_2023_tiger_nips}, we leverage item description text to derive item textual representations, and then apply a residual quantization method to train the tokenizer.

\paratitle{Item Textual Representations.}  Given the impressive language modeling ability of pretrained language models (PLMs), we use them to encode item description text into textual representations. Formally, for an item $i$ with text $V_i = \{v_1, v_2, \cdots, v_c\}$, its text embedding is defined as:
\begin{align}
    \bm{z}_i = \text{PLM}(v_1, v_2, \cdots, v_c),
\end{align}
where $\bm{z}_i \in \mathbb{R}^{d_w}$ denotes the mean-pooled representation of the last hidden states.

\paratitle{Tokenize via Residual Quantization.}
\label{sec:residual_quant}
Given the encoded item representations, we employ a classic quantization method, Residual-Quantized Variational AutoEncoder (RQ-VAE)~\cite{neil_2022_rqvae_ieee}, to generate a sequence of discrete codewords (\ie Semantic IDs). RQ-VAE tokenizer consists of an encoder, a decoder, and $m$ separate codebooks, each with $K$ entries. During tokenization, the model performs multi-level quantization over the latent vector produced by the encoder. At each level, the closest codeword is selected from the corresponding codebook, and the residual is updated for the next round. This recursive process allows the model to incrementally refine the representation. Specifically, given the input embedding $\bm{x}$, the encoder first produce a vector $\bm{z} = \mathcal{E}(\bm{x})$. The residual is initialized as $\bm{h}_1 = \bm{z}$. Then, for each level $d = 1, \ldots, m$, the residual is quantized using the corresponding codebook $\mathcal{C}_d = \{\bm{e}_k\}_{k=1}^K$. The nearest codeword is selected by $c_d = \arg\min_k \| \bm{h}_d - \bm{e}_k \|$. We then update the residual as $\bm{h}_{d+1} = \bm{h}_d - \bm{e}_{c_d}$ and repeat this process for $m$ levels. The final output is a sequence of $m$ discrete tokens $(c_1, \ldots, c_m)$, representing the Semantic ID. These tokens are then passed to the decoder to reconstruct the input $\bm{x}$. The optimization process can be formalized as:
\begin{align}
    \mathcal{L}_\text{RQ-VAE}(\bm{x}) &= \mathcal{L}_{\text{Recon}} + \mathcal{L}_{\text{Quant}},\\
    \mathcal{L}_{\text{Quant}} &= \sum_{d=1}^{m} \|\text{sg}[\bm{h}_d] - \bm{e}_{c_d}\|^2 + \beta \|\bm{h}_d - \text{sg}[\bm{e}_{c_d}]\|^2,\\
    \mathcal{L}_{\text{Recon}} &= \|\bm{x} - \hat{\bm{x}}\|^2,\\
    \hat{\bm{x}} &= \mathcal{D}(\sum_{d=1}^{m} \bm{e}_{c_d}),
\end{align}
where $\hat{\bm{x}}$ is the decoder’s output, and $\text{sg}$ denotes the stop-gradient operator.


\subsubsection{Backbone Model and Training} 
In this work, we adopt an encoder-decoder Transformer architecture (similar to T5~\cite{Raffel_2020_t5_jmlr}) as the backbone. Given the tokenized sequence of a user's historical interactions $S_u \rightarrow c_{1:p}$ and their target item $i_{n+1} \rightarrow C_{i_{n+1}} = c_{p+1:p+m+1}$, the model is trained using a sequence-to-sequence paradigm. Specifically, the encoder is trained to encode the user's historical interactions to capture implicit preferences, while the decoder autoregressively generates user target behavior.  Intuitively, this setup enables the model to make predictions token by token. However, under this setup, the decoder is only trained to generate target item tokens, resulting in limited output length (\ie tokenizer length $m$). This inevitably restricts the model’s ability to capture long-range dependencies and may hinder its slow thinking capacity.  To solve the above issue, we adopt a simple data synthesis strategy. Specifically, given the textual representation of the target item, we retrieve several items with similar representations and concatenate their corresponding tokens with the target item's tokens, thereby synthesizing a longer target response $Y$:
\begin{align}
\label{eqn:illustrate}
    Y = [\underbrace{c^\prime_1,\ldots, c_q^\prime}_{\text {tokens of retrieved items}};\quad \underbrace{c_{p+1}, \ldots, c_{p+m+1}}_{\text{tokens of target item}}],
\end{align}
where $q$ denotes the total number of tokens from the retrieved similar items. Then, the pretraining loss can be formalized as:
\begin{align}
    \label{eqn:loss_pretrain}
    \mathcal{L}_{\text{Pretrain}} &= - \sum_{t = 1}^{q+m}\log P_{\theta} (y_t | y_{< t}, c_{1:p}), \\
    P_{\theta} (y_t | y_{< t}, x) &= \frac{\exp(p(y_t)/\tau)}{\Sigma_{c \in \mathbb{C}} \exp(p(c)/\tau)},
\end{align}
where $\tau$ is the temperature, $\theta$ is the parameter of our backbone model.

\subsection{Supervised Fine-tuning for Adopting Slow Thinking Paradigm}
\label{sec:sft}
Based on the above pretraining method, our model can capture diverse user behavior patterns and support long-range modeling. However, its reasoning patterns remain inflexible (generating item tokens with similar representations to target items) and fall short of exhibiting intricate slow thinking patterns. In this section, we delve deeper into designing an effective reasoning process that encourages the model to perform sophisticated reasoning inference and provide more personalized recommendations. In what follows, we first present the slow thinking data collection process, and then describe the related training strategies.

\subsubsection{Iterative Inference for Data Collection}
\label{sec:iterative_inference}
\begin{figure*}[t]
	\centering
	\includegraphics[width=0.8\linewidth]{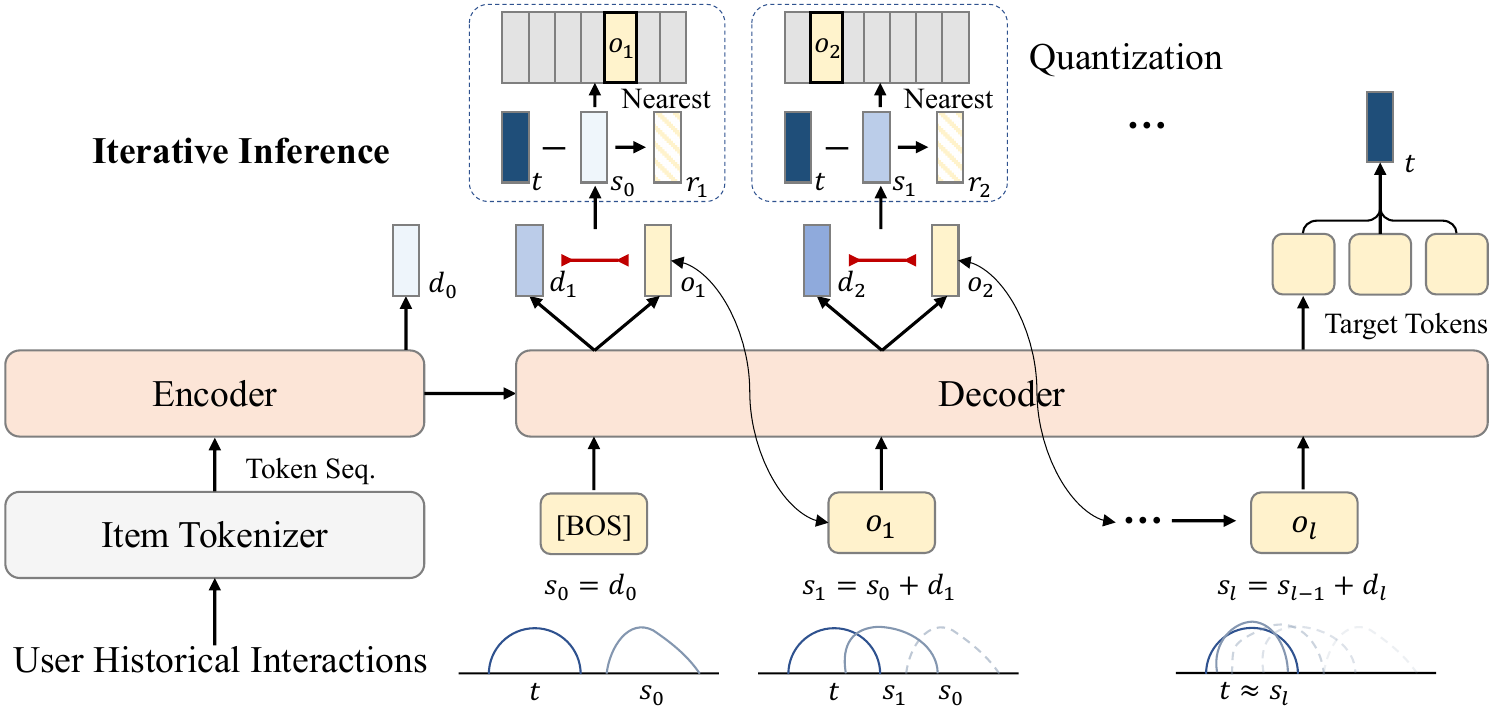}
        \caption{The iterative inference process of STREAM-Rec. We introduce it to collect intricate reasoning data for achieving slow thinking.    }
	\label{fig:sft}
\end{figure*}

As described in Section~\ref{sec:preliminary}, traditional sequential recommenders typically generate recommendations by calculating the matching degree between user behavior representations and candidate item representations. However, this method heavily relies on the model’s ability to accurately capture both user interests and item characteristics. In practice, smaller sequential models often struggle with such precise modeling through \emph{one-step} inference. To address this, we introduce an \emph{iterative} inference strategy that allows the model to progressively refine its predictions. Especially, the model is trained to estimate and predict the residual between its current prediction and the ground-truth representation of target item. The residual can serve as a corrective update, guiding the model toward the golden label. By recursively appending the residual vector to the decoder input and prompting the model to infer the next residual, our approach allows for step-by-step refinement until the output converges to the desired user behavior.
These intermediate residual data are then collected to train the model for achieving multi-step reasoning. To further clarify our iterative inference mechanism, it can also be viewed as an extension of the residual quantization algorithm for collecting intermediate reasoning data (see Section~\ref{sec:residual_quant} for a detailed explanation of Residual Quantization). The iterative inference process is illustrated in Figure~\ref{fig:sft}.

Specifically, given the tokenized user historical interaction sequence $S_u \rightarrow c_{1:p}$ and the target item $i_{n+1} \rightarrow C_{i_{n+1}}$, our encoder first encodes the historical token sequence into an initial state $\bm{s}_0 = \bm{d}_0$, representing an initial estimate of user preferences. The target item representation $\bm{t}$ is obtained by mean pooling over its token representations. The initial residual is defined as $\bm{r}_1 = \textit{MLP}(\bm{t} - \bm{s}_0)$, which means the gap between model prediction and golden label. 
Our objective is to iteratively reduce this gap and progressively refine the predictions closer to $\bm{t}$. To achieve this, at each step, the residual is incorporated into the input, and the state is updated accordingly to infer the next discrepancy. This iterative refinement process can be formalized as follows:


\textbf{Step 1. Pseudo-Label Generation:}
At iteration $i$ $(1\leq i \leq l)$, we compute the current residual as $\bm{r}_i = \textit{MLP}(\bm{t} - \bm{s}_{i-1})$, where $\bm{s}_{i-1}$ denotes the states from the $(i\!-\!1)$-th step. We then generate a pseudo-label token $o_i$ by selecting the token from the vocabulary $\mathbb{C}$, whose embedding best approximates the residual:
\begin{align}
o_i = \arg\min_{c_k \in \mathbb{C}} \| \bm{c}_k - \bm{r}_i \|,
\end{align}
where this pseudo-label $o_i$ represents the ideal corrective update at step \( i \), and it will be appended to the decoder input sequence to predict the updated decoder hidden state for the next inference.

\textbf{Step 2. State Update:} Given the predicted decoder hidden states $d_i$ in the $i$-th steps, we update the overall state by incorporating the current contribution as $\bm{s}_i = \textit{MLP}(\sum_{j=0}^{i}\bm{d}_j)$. Based on this, the new residual is then computed as $\bm{r}_{i+1} = \textit{MLP}(\bm{t} - \bm{s}_i)$. By repeating these two steps for  \( l \) iterations, we can obtain a sequence of $l$ tokens that represents an intermediate reasoning process. We then concatenate them with the target item tokens $C_{i_{n+1}}=\{c_{p+1}, \ldots, c_{p+m+1}\}$ to form the complete label sequence as follows:
\begin{align}
Y = [Y_\text{think} ; Y_\text{target}] =  [o_1, o_2, \ldots, o_l; c_{p+1}, \ldots, c_{p+m+1}].
\end{align}

\subsubsection{Training Strategies}
\label{sec:sft_training_strategies}
Based on the iterative inference algorithm described above, we can generate numerous potential thought processes. In this section, we further explain how to optimize the model to follow the above algorithm and produce more meaningful reasoning paths.

\textbf{State-Target Contrastive Loss.}
As mentioned before, during the iterative inference process, our goal is to progressively refine the model’s predictions to better align with target item representations. To facilitate this, we propose the state-target contrastive task. Specifically, given the predicted overall states at each timestamp, we treat the corresponding target item representations as positive samples, and use in-batch data as negative samples to conduct contrastive learning as follows:
\begin{align}
    \mathcal{L}_{\text{State}} = - \sum_{j=1}^{l} w_j \sum_{i=1}^B \log\frac{\exp(\bm{s}_{i,j} \cdot \bm{t_i})}{\sum_{j^\prime}^{B} \exp (\bm{s}_{i,j} \cdot \bm{t}_{j^\prime})},
\end{align}
where $\bm{s}_{i,j}$ denotes the state of the $i$-th sample at the $j$-th inference step, and $\bm{t}_i$ represents the target representation of the $i$-th sample. Note that the notation here slightly differs from that used in Section~\ref{sec:iterative_inference}, with an added dimension for clarity. $B$ denotes the batch size. $w_j$ is a weight coefficient that increases with the number of reasoning steps, reflecting our goal for the model’s predicted states to progressively converge toward golden target representations.

\textbf{Quantization Loss.} During the iterative process, we quantize the model-predicted residual to descrete tokens in the vocabulary $\mathbb{C}$, thereby generating a pseudo-label. To improve consistency, we introduce a quantization loss that aligns the residual representation with the corresponding quantized token:
\begin{align}
    \mathcal{L}_{\text{Quant}} = - \sum_{i=1}^B \sum_{j=1}^{l}  |\text{sg}[\bm{r}_{i,j}] - \bm{o}_{i,j}|^2 + \beta |\bm{r}_{i,j} - \text{sg}[\bm{o}_{i,j}]|^2,
\end{align}
where $\bm{o}$ is the representation of the quantized token $o$. During the overall optimization process, the straight-through estimator trick~\cite{bengio_2013_ste_trick_arxiv} is employed for gradient backpropagation.

\textbf{Supervised Fine-tuning Loss.} We employ the generated label sequence to fine-tune our model, enabling it to follow the slow thinking paradigm. The loss function $\mathcal{L}_{SFT}$ is similar to Eqn.~\ref{eqn:loss_pretrain}, maximizing the log likelihood. Notably, in contrast to the one-step recommendation, our goal is for the model to uncover deeper insights through a slower, more deliberate reasoning process. Therefore, to enhance the quality of thought processes, we deliberately train a \emph{direct} decoder as a reference model~($\pi_\text{ref}$), which shares the same encoder as our backbone model. This reference model is pretrained to perform one-step recommendations. We then compare the predictions of our model with those of the reference model using the DPO optimization strategy~\cite{Rafael_2023_dpo_nips}. This comparison enhances our model's ability to analyze and extract information throughout the slow thinking process.
Specifically, during DPO training, preferred samples are derived from user target items, while the dispreferred are sampled from items with similar representations to the target:
\begin{align}
\text{logit}^+ &= \log \pi_\theta(y^+|x, y_{\text{think}}) - \log \pi_{\text{ref}}(y^+|x), \\
\text{logit}^- &= \log \pi_\theta(y^-|x, y_{\text{think}}) - \log \pi_{\text{ref}}(y^-|x), \\
\mathcal{L}_{\text{DPO}} &= - \log \left(
    \frac{ \exp(\beta \cdot \text{logit}^+) }
         { \exp(\beta \cdot \text{logit}^+) + \exp(\beta \cdot \text{logit}^-) }
\right),
\end{align}
where $\text{logit}^+$ and $\text{logit}^-$ denote the log-probability differences between our model and the reference model for the preferred and dispreferred samples, respectively.

Overall, to jointly optimize our model, we combine the above loss functions as follows:
\begin{align}
    \mathcal{L} = \mathcal{L}_{\text{SFT}} + \lambda \mathcal{L}_{\text{DPO}} + \mu \mathcal{L}_{\text{Quant}} + \delta \mathcal{L}_{\text{State}},
\end{align}
where $\lambda$, $\mu$, and $\delta$ are weight hyper-parameters.

Notably, in Stage 2, the data collection process (Section~\ref{sec:iterative_inference}) and the optimization process (Section~\ref{sec:sft_training_strategies}) are carried out in a staggered manner. Specifically, once a substantial amount of reasoning data has been collected, we perform optimization over a fixed number of epochs to help the model memorize relevant patterns. We then resume iterative inference using the updated model to gather more precise reasoning data. These two processes are repeated iteratively until convergence.



\subsection{Reinforcement Learning for Exploring Effective Reasoning Patterns}
\label{sec:rl}
During the SFT process, our model is trained to fit the annotated trajectories. However, this training paradigm heavily depends on large-scale training data and may lead the model to memorize fixed patterns, resulting in suboptimal performance~\cite{chu_2025_sft_memorize_rl_generalize_arxiv}. To address this, we introduce reinforcement learning to further optimize the fine-tuned model, encouraging it to learn from experience and explore diverse reasoning patterns. In what follows, we first describe the employed reinforcement learning algorithm, and then present the reward design.

\subsubsection{Reinforcement Learning Algorighm} 
In this work, we employ the Group Relative Policy Optimization (GRPO) algorithm~\cite{shao_2024_grpo_deepseekmath_arxiv}. Specifically, GRPO updates the current policy (\ie our backbone model \(\pi_\theta\)), by leveraging two sources of information: a reference policy \(\pi_{\theta_{\text{ref}}}\), which serves as a regularization anchor, and trajectory rollouts sampled from a previously deployed policy \(\pi_{\theta_{\text{old}}}\). Given  \(G\) rollouts:
$
\tau = \{y_i\}_{i=1}^G \sim \pi_{\theta_{\text{old}}}(\cdot \mid s_u)
$, GRPO avoids training an explicit value function or critic. Instead, it estimates advantages \(A_i\) directly from the collected trajectories and uses them to guide the policy update. 

The objective is to improve the current policy by maximizing the expected surrogate reward, while also constraining the policy shift via KL regularization with respect to the reference policy. The full objective is defined as:
\begin{align}
\mathcal{J}(\theta) &= \mathbb{E}_{s_u \sim D, \{y_i\}_{i=1}^G \sim \pi_{\theta_{\text{old}}}(\cdot \mid s_u)} \nonumber \\
&\left[ 
\frac{1}{G} \sum_{i=1}^G 
\left( 
\min\left( 
\frac{\pi_\theta(y_i \mid s_u)}{\pi_{\theta_{\text{old}}}(y_i \mid s_u)} A_i,\,
\text{clip}\left( 
\frac{\pi_\theta(y_i \mid s_u)}{\pi_{\theta_{\text{old}}}(y_i \mid s_u)}, 
1 - \epsilon, 1 + \epsilon 
\right) A_i 
\right) 
- \beta \, \mathbb{D}_{\text{KL}}(\pi_\theta \| \pi_{\theta_{\text{ref}}}) 
\right) 
\right].
\end{align}

\subsubsection{Reward Design}
To effectively apply RL for optimizing the proposed slow-thinking recommender, we introduce several reward functions tailored to recommendation performance and slow thinking behavior.

\textbf{Format Reward.} As mentioned in Section~\ref{sec:preliminary}, the generative recommendation paradigm requires parsing the predicted item from the generated tokens. To improve this, we apply a penalty to generations that result in parsing errors:
\begin{align}
    \gamma_\text{Format} = \begin{cases}
0 & \text{if format is correct} \\
-1, & \text{otherwise}
\end{cases}
\end{align}

\textbf{Recommendation Exact Match Reward.} To encourage accurate recommendations, we apply a reward based on the exact match degree between model generated tokens and target item tokens. Specifically, given the generated tokens $\{\hat{c}_1, \ldots, \hat{c}_m\}$ and target item tokens $\{c_1, \ldots, c_m\}$, the reward function evaluates the number of consecutively matched tokens from the beginning of the sequence:
\begin{align}
\gamma_\text{EM} &= \sum_{j=1}^{m} \text{EMReward}(c_j, \hat{c}_j)\\
\text{EMReward}(c_j, \hat{c}_j) &= 
\begin{cases}
1, & \text{if } c_j = \hat{c}_j \text{ and } c_k = \hat{c}_k \ \forall k < j \\
0, & \text{otherwise}
\end{cases},
\end{align}

\textbf{Recommendation Similarity Reward.} Although the proposed exact match reward offers precise feedback, its signal is relatively rigid and may result in sparse feedback. To address this, we further compute the similarity between model predicted states and the target item representations, providing more fine-grained and softer feedback. Specifically, for a batch of samples, we compute the mean pooling of the model decoder’s hidden states $\bm{d} \in \mathbb{R}^{{B \times d}}$ and the target item token representations $\bm{T} \in \mathbb{R}^{{B \times d}}$, where $B$ denotes the batch size. We then compute the similarity between these representations to obtain a similarity matrix $S \in \mathcal{R}^{B \times B}$. For each sample $i$, its reward is defined as the proportion of samples in the batch whose similarity score is less than its correct match $\bm{S}_{i,i}$:
\begin{align}
    \gamma_{\text{Similarity}} &= 
    \begin{cases}
        0.5 & \text{if } g_i \geq 0.99 \\
        0.1 & \text{elif } g_i \geq 0.95 \\
        0.05 & \text{elif } g_i \geq 0.50 \\
        -0.1 & \text{otherwise}
    \end{cases} \\
    g_i &= \frac{1}{B} \sum_{j=1}^{B} \mathbb{I}(S_{i,j} \leq S_{i,i}),
\end{align}

\textbf{Thinking Likelihood Reward.} The above reward functions emphasize distinguishing the quality of recommendation predictions, here, we evaluate the quality of the generated thought process. Specifically, given the generated intermediate reasoning steps, we calculate the model’s likelihood score for generating the target item token based on this reasoning data. Furthermore, we compute the likelihood score of the direct recommendation decoder (see Section~\ref{sec:sft_training_strategies}) generating the target item in a one-step way, which serves as a regularization term:
\begin{align}
    \gamma_{\text{Likelihood}} = \text{LogProb}_{\theta} (Y_\text{target} | Y_\text{think}, s_u) - \text{LogProb}_{\text{ref}} (Y_\text{target} |  s_u) 
\end{align}
where $\text{LogPorb}(\cdot)$ is the logarithm of a probability value.

\textbf{Thinking Ranking Reward.} In addition to calculating the absolute likelihood score, we also consider the relative ranking of likelihood scores to provide more diverse feedback. Specifically, we sample $K$ hard negative examples whose textual representations are similar to the target item. 
We then compute the likelihood score for each of these samples based on the generated reasoning process. A reward is assigned if the likelihood of the positive sample ranks higher than the negatives; otherwise, a penalty is applied. Formally, let $p_i$ denote the rank of the positive sample in the $i$-th instance (with $p_i = 0$ indicating the highest rank), the ranking reward can be formalized as follows:
\begin{align}
    \gamma_{\text{Ranking}} =
\begin{cases}
0.2, & \text{if } p_i = 1 \\
0.1, & \text{elif } 1 < p_i < 0.1K \\
0.05, & \text{elif } 0.1K \leq p_i < 0.2K \\
-0.1, & \text{elif } p_i \geq 0.5K \\
\end{cases}
\end{align}

The overall reward for RL training is the sum of the above rewards.

\section{Experiment}

\subsection{Experiment Setup}
\subsubsection{Dataset}
To evaluate the performance of the proposed approach, we conduct experiments on two subsets of Amazon Review datasets~\cite{AmazonReview}: ``Musical Instruments'', and ``Industrial and Scientific''. Following previous work~\cite{S3Rec}, we keep the five-core datasets
and filter users and items with fewer than five interactions. Then, we group the interactions by users and sort them
by timestamp ascendingly. The maximum sequence length is set to 20. The statistics of datasets are summarized in Table~\ref{tab:data_statistics}.

\begin{table}[]
\centering
\caption{Statistics of the preprocessed datasets. ``Avg.$n$'' is the average length of behavioral sequences.}
\label{tab:data_statistics}
\begin{tabular}{lrrrrr}
\toprule
 \textbf{Datasets} & \textbf{\#Sequences} & \textbf{\#Items}  & \textbf{\#Actions} & \textbf{Avg.$n$} & \textbf{Sparsity} \\
\midrule
\textbf{Instrument} & 57,349  & 24,587 & 511,836 &7.37  & 99.96\% \\
\textbf{Scientific} & 50,985 & 25,848  & 412,947  & 8.10 & 99.97\% \\
\bottomrule
\end{tabular}
\end{table}

\subsubsection{Baseline Models}
We compare our proposed method against several representative sequential recommendation models:

(1) \textbf{Caser}~\cite{caser} is a CNN-based sequential recommendation model that captures patterns in user behavior sequences using both horizontal and vertical convolutions.

(2) \textbf{GRU4Rec}~\cite{gru4rec} utilizes the standard GRU architecture to encode user interaction sequences.

(3) \textbf{SASRec}~\cite{SASRec} is a classic sequential recommendation model based on a transformer encoder. It uses a multi-head self-attention mechanism to capture sequential patterns in user behavior.

(4) \textbf{BERT4Rec}~\cite{BERT4Rec} is a bidirectional self-attention model that adopts a cloze-style objective to improve sequence representation learning.

(5) \textbf{TIGER}~\cite{Rajput_2023_tiger_nips} is a generative recommendation model that constructs semantic codes from item text embeddings and leverages the T5 architecture to generate recommendations.

\subsubsection{Evaluation Settings}
To assess the performance on the next item prediction task, we adopt two widely used metrics: \ie HR@$N$ and NDCG@$N$ (in short, R@$N$ and N@$N$), where $N$ is set to 5 and 10. Following previous work~\cite{S3Rec}, we use the leave-one-out evaluation strategy. For each user interaction sequence, the last item is held out as the test instance, the second-to-last item is used for validation, and the remaining interactions are used for training. During evaluation, we rank the target item of each sequence against all other items in the test set and report the average metric score across all test samples.

\subsubsection{Implementation Details}
We implement our model and all baseline methods in PyTorch. To ensure a fair comparison, we use the AdamW optimizer for both our model and the baselines, and perform the hyperparameter search. We instantiate the T5 architecture with 4 encoder and 4 decoder layers. The hidden size is set to 256, and the FFN dimension is set to 1024. Each layer includes 4 self-attention heads, each with a dimensionality of 64. The tokenizer length $m$ is set to 4. In Stage 1, we retrieve 2 items with representations similar to the target items. The learning rate is set to 0.001.
In Stage 2, during iterative inference to collect reasoning data, we set $l = 6$ for the Instruments dataset and $l = 4$ for the Scientific dataset. The learning rate for our model and the baselines is tuned from $\{0.001, 0.003, 0.005\}$.
In Stage 3, we retrieve 50 similar items to serve as negative samples when computing the thinking ranking rewards. The learning rate in this stage is set to 0.00001. All the experiments are conducted on 8 NVIDIA 3090 GPUs. We use the early stopping strategy with a patience of 10 epochs to prevent overfitting, and NDCG@10 is used as the indicator metric.

\begin{table}[t]
   \caption{Overall performance of different recommendation methods. The best and the second-best performance methods are denoted in bold and underlined fonts, respectively. “Improv.” denotes the relative improvement ratios over the TIGER baseline, which is a ``fast thinking'' generative recommender and represents the simplified version of our method (excluding the reasoning process). ``PT'', ``SFT'', and ``RL'' refer to the model obtained at three training stages.}
    \centering
    \small
    \begin{adjustbox}{width=\linewidth,center}
    \begin{tabular}{lrrrrrrrrr}
        \toprule
        \multirow{2.5}[0]{*}{Method} & \multicolumn{4}{c}{\textbf{Instruments}} & \multicolumn{4}{c}{\textbf{Scientific}} \\
        \cmidrule(lr){2-5} \cmidrule(lr){6-9}
        & R@5 & R@10 & N@5 & N@10 & R@5 & R@10 & N@5 & N@10 \\
        \midrule
        \multicolumn{9}{c}{\textbf{Traditional Methods}} \\
        \midrule
        Caser       & 0.0151 & 0.0256 & 0.0094 & 0.0128 & 0.0098 & 0.0167 & 0.0062 & 0.0085 \\
        GRU4Rec     & 0.0230 & 0.0353 & 0.0152 & 0.0191 & 0.0166 & 0.0251 & 0.0116 & 0.0143 \\
        SASRec      & 0.0352 & 0.0556 & 0.0220 & 0.0286 & \underline{0.0255} & 0.0384 & 0.0158 & 0.0200 \\
        BERT4Rec    & 0.0257 & 0.0414 & 0.0162 & 0.0212 & 0.0170 & 0.0277 & 0.0106 & 0.0141 \\
        \midrule
        \multicolumn{9}{c}{\textbf{Generative Methods}} \\
        \midrule
        TIGER                    & 0.0333 & 0.0532 & 0.0215 & 0.0279 & 0.0240 & 0.0382 & 0.0153 & 0.0199 \\
        \cmidrule(lr){1-9}
        STREAM-Rec$_\text{PT}$   & 0.0218 & 0.0316 & 0.0139 & 0.0170 & 0.0196 & 0.0315 & 0.0124 & 0.0162 \\
        Improv.                  & --     & --     & --     & --     & --     & --     & --     & --     \\
        \cmidrule(lr){1-9}
        STREAM-Rec$_\text{SFT}$  & \underline{0.0367} & \underline{0.0562} & \underline{0.0244} & \underline{0.0307} & 0.0249 & \underline{0.0390} & \underline{0.0159} & \underline{0.0204} \\
        Improv.                  & +10.21\% & +5.64\% & +13.49\% & +10.04\% & +3.75\% & +2.09\% & +3.92\% & +2.51\% \\
        \cmidrule(lr){1-9}
        STREAM-Rec$_\text{RL}$   & \textbf{0.0398} & \textbf{0.0608} & \textbf{0.0261} & \textbf{0.0329} & \textbf{0.0264} & \textbf{0.0424} & \textbf{0.0169} & \textbf{0.0221} \\
        Improv.                  & +19.52\% & +14.29\% & +21.40\% & +17.92\% & +10.00\% & +10.99\% & +10.46\% & +11.06\% \\
        \bottomrule
    \end{tabular}
    \end{adjustbox}
    \label{tab:overall_result}
\end{table}

\subsection{Overall Performance}
We compare our approach with the baseline methods on two datasets and present the results in Table~\ref{tab:overall_result}. We can find that:

(1) Traditional sequential recommendation methods, such as SASRec, BERT4Rec, and GRU4Rec do not perform well. It is possibly because they have to capture user preference, item characteristics, and their matching relations in a one-step inference. However, these traditional small models are incapable of achieving such precise modeling. For generative sequential recommendation method, \ie TIGER, achieves better results than traditional sequential recommender in most cases, this result indicates the effectiveness of generative modeling. Nevertheless, existing generative recommendation methods still have to directly inferring user target behavior based on their historical interactions. 

(2) Our proposed \textit{STREAM-Rec} follows a three-stage training process. As shown by the results, model experiences a noticeable performance drop after the pretraining stage. This decline is primarily attributed to the low-quality reasoning processes synthesized during pretraining.  Notably, we retrieve a set of items with representations similar to the target item and concatenate their tokens with the target item's tokens to generate pretraining data.  However, these low-quality reasoning processes do not effectively guide the model in progressively inferring user target behavior.   Nevertheless, pretraining on such long sequences does help the model capture long-range dependencies.

(3) In Stage 2, we leverage the pretrained recommender to perform iterative inference, progressively refining its predictions using target item representations. This process allows us to synthesize large-scale, high-quality reasoning data. As shown, fine-tuning on this data enhances the model’s slow thinking ability, enabling it to outperform traditional one-step generative recommenders. Furthermore, by conducting reinforcement learning, our model can flexibly explore more reasoning patterns, leading to even better performance.

\section{Related Work}

\paratitle{Sequential Recommender Systems.}
Due to the dynamic nature of user preferences, researchers have proposed sequential recommendation methods that aim to predict future items by uncovering hidden patterns in user behavior sequences. To this end, various advanced architectures such as RNNs~\cite{gru4rec}, CNNs~\cite{caser, fpmc}, and Transformers~\cite{SASRec, zhang_2024_aurisrec_emnlp} have been introduced to more effectively model user preferences.
More recently, several studies have explored generative approaches for recommendation, where user behavior sequences are tokenized into sequences of discrete tokens.   While these methods have shown promising performance, they often follow a common recommendation paradigm: modeling user preferences by encoding historical interactions and then \textbf{directly} computing the \textbf{matching} score between the encoded preferences and the representation of a candidate item (in traditional methods) or token (in generative methods). However, this single-step recommendation paradigm may constrain the model’s ability to generate more thoughtful and deliberate recommendations.

\paratitle{Slow Thinking.}
Recently, slow thinking reasoning systems such as OpenAI o1 and DeepSeek R1 have shown impressive performance on complex reasoning tasks. These systems typically generate valuable thinking processes before providing a response, enabling them to produce more comprehensive, and well-justified solutions~\cite{chen_2025_still_v1_arxiv, zhang_2025_test_time_scaling_survey_arxiv, zhao_2023_llm_survey_arxiv}. Inspired by these advances,  several approaches have been proposed to incorporate slow thinking mechanisms into recommender systems, aiming to provide more personalized recommendations. For instance, ReaRec~\cite{tang_2025_rearec_arxiv} improves user representations by autoregressively feeding the last hidden state of the sequence into the sequential recommender. However, this approach adopts an ``implicit'' multi-step reasoning paradigm. In contrast, our method employs iterative inference and reinforcement learning to perform explicit reasoning steps.
Moreover, Rec-R1~\cite{lin_2025_recr1_arxiv} addresses the LLM4Rec task by promoting the reasoning capacities of LLMs to generate more suitable recommendations. Unlike Rec-R1, our approach delves into the reasoning patterns of smaller generative recommenders and demonstrates their efficacy in performing slow thinking.

\section{Conclusion and Future Work}
In this work, we proposed \textit{STREAM-Rec}, a generative recommender that performs slow thinking to provide tailored recommendations. Our approach creatively stimulates the reasoning capabilities of traditional recommenders through a three-stage training framework. Specifically, in stage 1, we pretrain our model on numerous user behavior data, enabling it to capture complex behavioral patterns and model long-range dependencies. In stage 2, we propose an iterative inference algorithm to collect sophisticated reasoning traces, by progressively refining the model's predictions toward user target behavior representations. The model is then fine-tuned on this data to adopt the slow thinking paradigm. To further improve generalization, in Stage 3, we design several reward functions and apply reinforcement learning to encourage the model to explore more effective reasoning patterns. Based on these methods, our model can generate a meaningful thinking process before providing a final recommendation, thereby significantly enhancing its overall performance.

It is worth emphasizing that, although our approach has achieved promising preliminary results, there remains substantial room for further improvement:

\textbf{Test-time Scaling.} This paper offers an initial investigation into the slow thinking capabilities of traditional recommendation models. In our current framework, the length of the reasoning process is fixed. In the domain of large language models, some studies have demonstrated that longer reasoning paths can enhance performance~\cite{chen_2025_still_v1_arxiv, zhang_2025_test_time_scaling_survey_arxiv}, while others have highlighted the risk of ``overthinking'' when the reasoning process becomes excessively long~\cite{luo_2025_overthink_arxiv, chen_2024_overthink_arxiv}. In the context of recommendation systems, it remains an open question: \emph{Does a longer reasoning process lead to better recommendation accuracy?}

\textbf{Training-time Scaling.} Due to computational constraints, our experiments are conducted on relatively small-scale recommendation models. In future work, we plan to explore whether scaling up the model size and training data can further improve its reasoning capabilities.

\textbf{Golden Reasoning Paradigm.} We introduce an iterative inference algorithm that progressively estimates and reduces the residual between model predictions and target user behavior representations, to annotate potential reasoning trajectories. While this strategy improves reasoning ability and yields strong empirical performance, it inevitably relies on manual priors. To mitigate this, we incorporate reinforcement learning techniques for the fine-tuned model to discover more effective reasoning patterns. However, we empirically find that applying reinforcement learning directly to a cold-start model often results in suboptimal outcomes. A more rigorous theoretical analysis is needed to identify the optimal form of reasoning.

\textbf{Efficiency Optimization.} While our method can activate the slow-thinking potential of traditional recommendation models, the multi-step reasoning paradigm introduces greater computational overhead compared to conventional single-step approaches. Future research will explore advanced optimization techniques, such as quantization~\cite{zhou_2024_llm_quant_survey_arxiv} and distillation~\cite{cheng_2024_compress_cot_arxiv}, to improve inference efficiency.

\bibliographystyle{plain}
{\small
\bibliography{reference}}
\end{document}